\documentclass[12pt,preprint,showpacs]{revtex4}
\usepackage{amssymb}
\usepackage{amsmath}
\usepackage{graphicx}
\usepackage{mathrsfs}
\usepackage{color}
\usepackage[colorlinks]{hyperref}
\usepackage{natbib}

\begin{document}

\title{Theoretical investigation of thermostability of incompressible channels in quantum Hall states}
\author{Tao Yang$^{1,2}$, and Keith A. Benedict$^1$}
\affiliation{$^1$School of Physics and Astronomy, University of Nottingham, Nottingham, NG7 2RD, United Kingdom\\
$^2$Institute of Modern Physics, Northwest University, Xian, 710069, P. R. China}

\begin{abstract}
In this work we use self-consistent method considering a two dimensional electron gas system in the integer quantum Hall regime, to calculate the temperature induced decay of incompressible stripes. There are two types of incompressible strips which can form in a Hall bar system by varying the electron density or magnetic field. We observe that the way of collapse of incompressible strips strongly depends on the type of them. With increasing temperature a bulk incompressible strip will decay from the middle and separate into two edge channels by a density ramp, while an incompressible edge channel reduces its size from both sides until vanishes.
\end{abstract}


\maketitle

\section{Introduction}

The formation of alternating strips of compressible and incompressible strips \cite{PRB.46.4026,PRB.47.12605,physicab.298.155} in a gate-confined two-dimensional electron gas (2DEG) is important for understanding the mechanism of many phenomena in the quantum Hall regime, such as the transport properties of the edge states\cite{physicae.3.30,physicae.20.43,physicae.40.1217,physicae.40.1232}, heat transport in quantum Hall effect (QHE) samples\cite{arXiv.1202.6681v1}, spatial distribution of local electron temperature\cite{physicae.44.1367,PRB.73.045333}, quantum Hall breakdown (QHBD)\cite{physicae.4.79,cage,PRL.51.1374}, topologically protected states in the fractional QHE\cite{RMP.80.1083} and in novel topological states at zero magnetic field\cite{Science.318.766}. Recent experimental investigation of the microscopic origin using local probe techniques\cite{PRL.101.256802,Nature.427.328,PRL.107.176809, NJOP.14.083015} makes it possible to image these strips. Therefore, further study of the structure of the strips is important for an accurate description of QHE and related topics.

The phenomenon is induced by the screening effect of a 2DEG, characterized by the density redistribution of the electrons in response to an imposed external potential in order to minimize the total energy of the system, which is closely related to the metal-insulator transition. The screened potential is then the sum of the external potential and the Coulomb potential of the redistributed electrons. When the density modulation of the 2DEG is smaller than the average density, the screen is linear and the external potential is greatly reduced. With decreasing electron density, or increasing fluctuations of the disorder potential, electrons begin to be depleted locally where the external potential is not screened, which means that the screening is nonlinear. In the presence of a strong perpendicular magnetic field a 2DEG shows unusual low-temperature screening properties \cite{PRB.38.4218,SSC.67.1019}, since the highly degenerate quantized energy levels, i.e. Landau levels (LLs), lead to a strong variation of the density of states (DOS) with varying strength of the magnetic field. Strong localization can happen due to the Landau gaps in the energy spectrum \cite{RMP.81.109}. When the chemical potential sits between two successive LLs, there is no screening if the external potential is not strong enough and the temperature is very low because the energy supplied by the disorder potential and the thermal fluctuation is not strong enough to excite electrons to higher
LLs. Therefore, an incompressible region is formed. If the chemical potential is pinned in a LL we expect a nearly perfect screening at a low temperature. This happens when the amplitude of the fluctuation of the disorder potential is not large enough to cause LLs to overlap, and then only the electrons in the partially filled level are free to adjust their density. In Ref.\cite{PRB.45.11354} a theory for transport that includes screening effect on the IQHE were presented. Self-consistent calculations of screening properties in a Hall bar system have been developed by Gerhardts \textit{et al.} \cite{PRB.50.7757,PRB.56.13519,PRB.68.125315,PRB.70.195335}. This provides us a good tool to study the stationary screening properties of a 2DEG.

The dependence of the widths and position of compressible and incompressible strips on the filling factor\cite{PRB.70.195335,NJOP.14.083015}, confinement potential\cite{PRB.46.4026,arXiv.1203.4985,arXiv.1210.8425} and imposed external current\cite{PRB.70.195335} has been widely studied both theoretically and experimentally.
In this paper, we investigate the thermostability of these states in a Hall bar system under the translational invariance consideration, which is closely related to the temperature induced QHBD.

\section{Self-consistent Calculations}
The 2DEG system in a perpendicular magnetic field $\textbf{B}$ is actually a Hall bar system. The energy spectrum of 2DEG is split into different LLs.  We denote the filling factor
\begin{equation}\label{0-7}
\nu(\textbf{r})\doteq\rho(\textbf{r})/n_B=2\pi l_B^2\rho(\textbf{r})
\end{equation}
as the number of filled Landau levels, where $\rho(\textbf{r})$ is the density distribution of electrons, $l_B=\sqrt{\hbar/eB}$ is the magnetic length and $n_B$ is the the state density of a LL. To get this result we have assumed that the magnetic field is strong enough to make complete spin polarization. To calculate the non-linear screening, we have to consider all the occupied energy levels (or the total density of electrons), and then calculate the electron density by employing the Thomas-Fermi approximation. Essentially, the screening effect has arisen because of the interaction between electrons. For a given density distribution, we will get the corresponding potential profile, and it in turn will change the density distribution until the electrostatic energy of the system is minimized. It is actually a self-consistent problem and can be described by the following equations:
\begin{eqnarray}
V(\textbf{r}) &=& V_{conf}( x ) + V_{int}( \textbf{r})\nonumber\label{sc2-1},\\
\rho( \textbf{r} ) &=& \int d\epsilon D(\epsilon) f( \epsilon + V(\textbf{r}) - \mu )\label{sc2-2},
\end{eqnarray}
where $\mu$ is the chemical potential, and $f(\epsilon)=1/(1+e^{\epsilon/k_BT})$ is the fermion distribution with $k_B$ being the Boltzmann¡¯s constant and $T$ being the temperature. 
The direct Coulomb potential is
\begin{equation}
V_{int}(\textbf{r})=\frac{e^2}{2\varepsilon}\int d^2r'\frac{\rho(\textbf{r})\rho(\textbf{r}')}{|\textbf{r}-\textbf{r}'|},
\end{equation}
and the confinement potential, which can be simply set as only a function of $y$ by supposing that the system has translational invariance in the longitudinal direction ($x$-direction), can be written as \cite{PRB.70.195335}
\begin{equation}
V_{conf}( y ) = -\frac{2\pi e^2}{\epsilon}\rho_{bg}\sqrt{d^2 - y^2}
\end{equation}
where $d$ is half of the sample width, $\rho_{bg}$ is the uniform background positive charge density.

The bare Landau DOS without considering the broadening effect of LLs by impurities is given by
\begin{equation}\label{dos2}
D(\epsilon)=\frac{1}{2\pi l_B^2}\sum_0^\infty\delta(\epsilon-\epsilon_n),
\end{equation}
where $\epsilon$ is the energy and $\epsilon_n$ is the energy of the $n_{th}$-LL.

\begin{figure}[t]
\begin{center}
\includegraphics[scale=0.45]{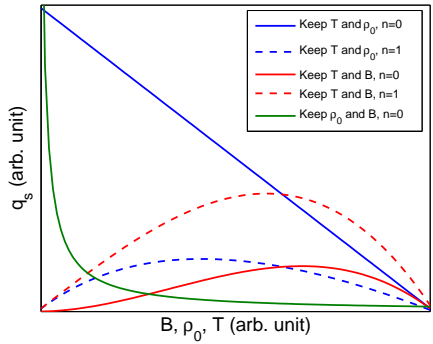}
\caption{The effective screening wave vector as a function of a magnetic field $B$ (blue lines), electron average density $\rho_0$ (red lines) and temperature $T$ (green line), by keeping other parameters fixed.}\label{fig-scr-length}
\end{center}
\end{figure}

One of the most important physical quantities for the screening effect is $q_s$, the effective screening wave number, which measures how effectively a given external potential is reduced due to electron interactions. If the chemical potential is pinned in a LL we expect a nearly perfect screening at a low temperature. This happens when there is no LL overlap, and then only the electrons in the partially filled level are free to adjust their density. For a non-integer filling factor which can be expressed by $\nu=n+\bar{\nu}_n$, where $n$ is the number of fully occupied LLs and and $0<\bar\nu_n< 1$ is the average filling factor of the partially occupied LL (highest LL), the effective screening wave vector for finite temperature and magnetic field is given as\cite{PRB.38.4218}
\begin{equation}\label{scr_length1}
q_s =\frac{\hbar\omega_c}{k_BT}\bar\nu_n(1-\bar\nu_n)\frac{1}{a_B},
\end{equation}
where $a_B = \epsilon\hbar^2/e^2m^*$ is the effective Bohr radius. There are three variables that can be freely adjusted in the above equation, the magnetic field $B$, the average density of electrons $\rho_0$, and the temperature $T$. The average filling factor of the highest occupied LL, $\bar\nu_n$, is a function of the magnetic field and the density of electrons. The change of $q_s$ with respect to the different parameters is given in Fig. \ref{fig-scr-length}. If we keep $\rho_0$ and $T$ constant but decrease $B$ the change in $q_s$ is decided by the function $\bar\nu_n(1-\bar\nu_n)/(n+\bar\nu_n)$. If $ n= 0$ (only the lowest LL is occupied) and the strength of the magnetic field is reduced, $q_s$ decreases linearly in response to the increasing $\bar\nu_n$ (solid blue line). If $n$ is not zero (more than one LLs are occupied), $q_s$ will increase to its maximal value at about $\bar\nu_n = 1/2$, and then decrease as $\bar\nu_n$ approaches 1 (dashed blue line). This means that nearly perfect screening at about $\bar\nu_n=1/2$ as reported in Ref.\cite{Nature.427.328}. For a filling factor higher than $1/2$ the screening is less effective. If we keep $B$ and $T$ constant and change $\rho_0$, the change in $q_s$ depends on the function $\bar\nu_n(1-\bar\nu_n)(n+\bar\nu_n)$. The filling factor $\bar\nu_n$ decreases with $\rho_0$. The screening in this case is similar to the case where only $B$ is changed and the occupation of LLs is larger than 1 (the red lines). The simplest case is where only $T$ is varied, and $B$ and $\rho_0$ are kept constant. Then $q_s$ decreases with increasing $T$ (the green line), which means that the screening is less effective at high temperatures, and the density modulation is accordingly smaller than the low temperature screening. We can see that $q_s$ decreases rapidly as the temperature rises (solid green line), while the effect of other variables is fairly mild. Thus, it is safe to assume that the screening becomes poor without considering the effect of the density modulations.

The basic idea of solving Eq.(\ref{sc2-2}) is that we start with a guess for the density profile, $\rho(\textbf{r})=\rho_0$ for example, calculate the Coulomb potential, solve Eq.(\ref{sc2-2}) to find better estimates for the density profile, and repeat the loop until the density profile ceases changing. We note that the self-consistent method is applicable for both linear and the nonlinear screening, and all kinds of external potentials in a realistic sample can be included by using this method, such as donor potential, which will not be considered in this paper.

\section{Incompressible bulk and edge channels}

\begin{figure}[t]
\begin{center}
\includegraphics[scale=0.75,bb=-10 250 778 500]{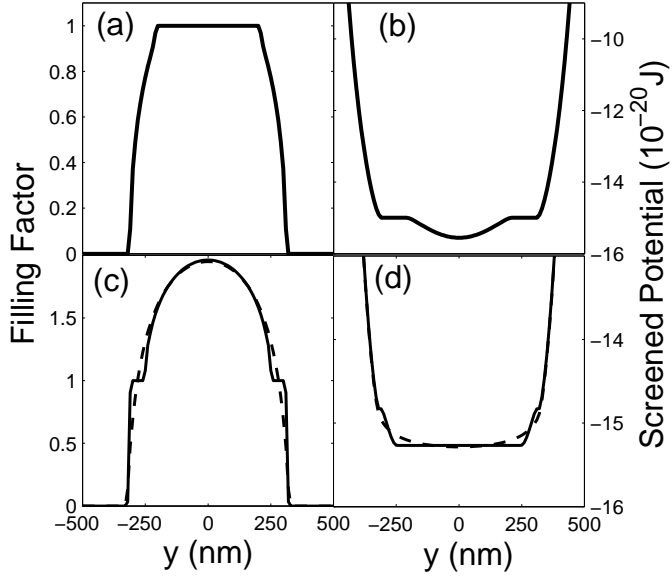}
\caption{Distribution of the filling factor and the screened potential of the confined 2DEG system without impurities. (a)(b) The average filling factor is $\bar\nu=0.55$. The bulk incompressible region with $\nu=1$ appears in the middle of the system. (c)(d) The average filling factor is $\bar\nu=1$. The dashed lines are the results in the absence of magnetic field, where we have used the Landau DOS to get the 'filling factor' for comparison in (c). }\label{bulk-incompressible}
\end{center}
\end{figure}

\begin{figure}[t]
\begin{center}
\includegraphics[scale=0.6]{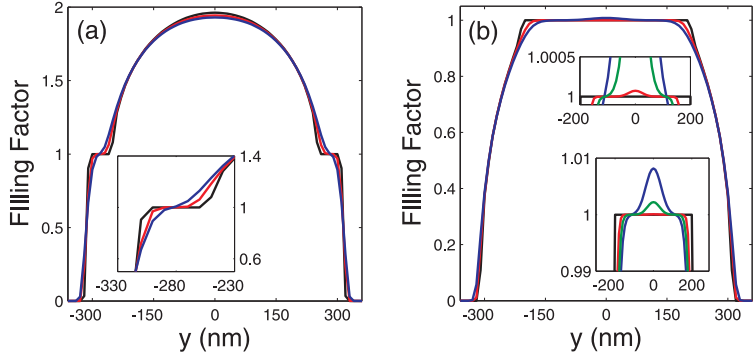}
\caption{Distribution of the filling factor for the average filling factor $\bar\nu=1$ (a) and $\nu=0.55$ for $k_BT/\epsilon_F=0.001$ (black line), 0.1 (red line), and 0.2 (blue line). The insets are zoom plots for the incompressible region. The extra green line in the insets of (b) corresponds to $k_BT/\epsilon_F=0.15$. At about $k_BT/\epsilon_F=0.2$, both edge channels and bulk channel collapse.}\label{step-decay}
\end{center}
\end{figure}

\begin{figure}[t]
\begin{center}
\includegraphics[scale=0.6]{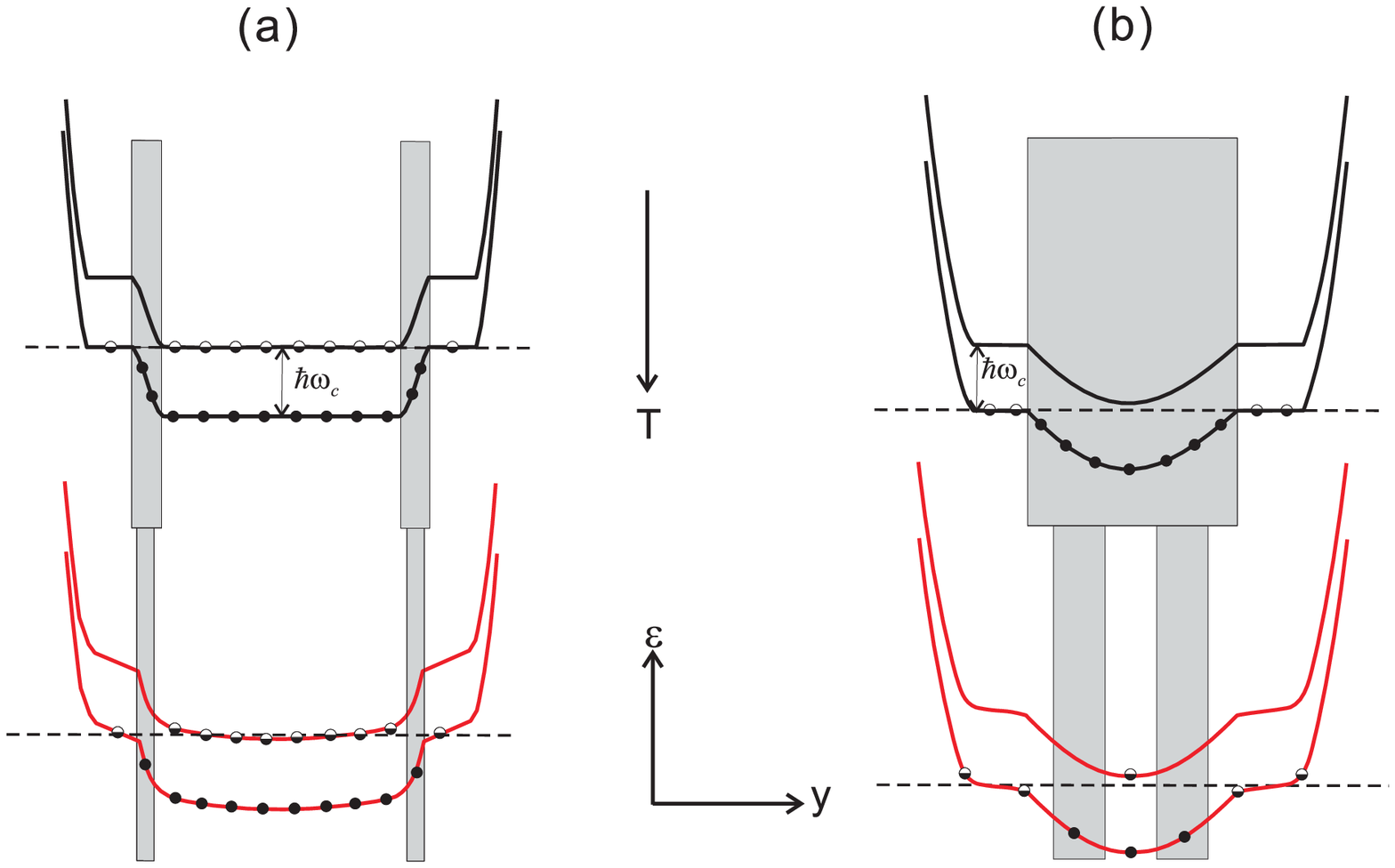}
\caption{Schematic diagrams of the occupation of LLs for the temperature induced breakdown of incompressible strips (shaded area) in a Hall bar system. (a) There are two incompressible edge channels at a low temperature. The width of these two channels decreases when the temperature $T$ is increased until they disappear. (b) There is only a bulk incompressible channel at a low temperature. When the temperature increases, this channel breaks into two edge channels from the centre. Then the edge channels will shrink with an increasing temperature. The solid black lines are LLs at a low temperature which are bent by the screened potential. The solid red lines are the LLs at a higher temperature. The dashed lines are the chemical potential of the system. The solid circles indicate that a LL is locally fully occupied (local filling factor is a integer), while the half-filled circles indicate that a LL is locally partially occupied. See the text for a fuller description.}\label{fig-thermal}
\end{center}
\end{figure}

By varying the magnetic field (and hence the average filling factor), the width and position of the incompressible region change accordingly. For the filling factor $\nu\leq1/2$, there are no incompressible strips. For the filling factor $1/2<\nu\leq1$, the system will undergo a transition from the bulk incompressible state to the incompressible edge state. We choose two typical average filling factors 0.55 and 1 for a sample with average density of electrons $\rho_0=4\times 10^{11}cm^{-2}$.

Figs.\ref{bulk-incompressible}(a) and \ref{bulk-incompressible}(b) shows the results for the situation where the average filling factor is 0.55. The bulk incompressible region with $\nu=1$ appears in the central part of the system, where the confinement potential is not screened. The confinement potential is only screened along the edges, where the screened potential is a constant. The results for $\bar\nu=1$ are shown in Fig.\ref{bulk-incompressible}(c) and \ref{bulk-incompressible}(d). The screening is better when the magnetic field is applied, as can been seen by comparing this with the results in the absence of the magnetic field (dashed lines in Fig.\ref{bulk-incompressible}). Two steps with local filling factor $\nu=1$ appear around the edge of the system, which means the formation of incompressible edge channels and no screening in these regions. The compressible regions are separated by the edge channels. The screened potential becomes nearly constant within the central compressible region (the region between the two plateaus in Fig.\ref{bulk-incompressible}(c)), which behaves like metal strip at constant potential. If we consider the spin degeneracy, there will be only incompressible regions with even local filling factors. A recent study \cite{arXiv.0906.3796}, based on a Hall bar design on a cleaved edge overgrown wafer by the help of a side gate, shows that the widths of the incompressible strips can be changed and even made to vanish when changing the edge potential probe. Such control of the edge potential implies peculiar transport results when considering the screening theory, which includes the direct Coulomb interaction explicitly. The underlying physics is intrinsically the same as the confinement potential we discussed here. We note that there can be a state with a mixture of a bulk incompressible channel and incompressible edge channels if the average filling factor is higher.

The effect of finite temperature on the decay of the incompressible strips is demonstrated in Fig.\ref{step-decay}. The values of $T$ are chosen to yield $k_BT/\epsilon_F=0.001,~0.1,$ and $0.2$ ($T=0.16,~15.87,~31.75K$). At zero temperature, the filling factor profile should be vertical on each side of the incompressible channels (the flat regions in this plot), as reported in Ref.\cite{PRB.46.4026}. The result of our simulation at $T=0K$ is close to this limiting case, but it has not yet been reached. The width of the strips decreases with increasing temperature, and there are no incompressible regions at $T=31K$. This agrees with the experiment in Ref.\cite{SSC.54.479} where the temperature is reported to be about $30K$. The pattern difference between the collapse of the edge channels and the bulk channel is due to the role of the confinement potential, and the respective mechanisms are shown schematically in Fig.\ref{fig-thermal}.

With an external confinement potential the LLs are bent and have the same shape as the screened potential. In Fig. \ref{fig-thermal}(a) the central part of the sample, where the screening is nearly perfect, has a fully occupied Lowest LL and a partially occupied second LL. This configuration leads to straight LLs. In the incompressible edge channels (the shaded area) the lowest LL is fully occupied and the second one is empty, leaving the confinement potential unscreened there. Beyond these channels, the filling factor decreases from 1 to 0 towards the boundaries of the sample, creating two steps with perfect screening. The chemical potential of the system is pinned in the second LL (dashed lines). %
When temperature is increased, the LLs are bent at the compressible regions and some electrons in the lowest LL are excited to the second LL through the edge channels where the potential is equal to the chemical potential. The density of electrons between the incompressible edge channels and the boundary of the sample reduces. The density of electrons in the region of the central compressible area close to the edge channels increases slightly because of the large bending of the LLs there (see Fig.\ref{step-decay}(a)). This makes the filling factor increase on the inner side of the edge channels and decrease on the outer side. These channels then shrink from the both sides as the temperature increases, and eventually vanish. The bending of the LLs in the centre of the compressible region is weak, and the filling factor here decreases as shown in Fig.\ref{step-decay}(a).

For a system with a bulk incompressible channel the situation is different. We will assume that only the lowest LL is occupied at low temperatures. In the central area the filing factor is 1 and there is no screening. Two steps appear close to the boundaries of the sample where the lowest LL is partially occupied and the screening is perfect. There is a small gap between the level of chemical potential of the system and the bottom of the second LL. The electrons are excited thermally to the second LL as the temperature increases, and they will occupy the states at the bottom of the second LL where the energy is lower. This generates a peak in the filling factor profile (see Fig. \ref{step-decay}(b)), where the bulk incompressible channel is divided into two incompressible channels as seen in Fig.\ref{fig-thermal}(b). With increasing temperature, more electrons are excited into the second LL, and the chemical potential increases accordingly. The central compressible channel expands to the boundaries of the sample. The two incompressible channels decay in the same way as the edge channels described above.

\section{conclusion}\label{sec-con}

In this paper, we investigated the formation of the bulk incompressible state and the incompressible edge state for a Hall bar system in the presence of a confinement potential and calculated the temperature induced decay of a bulk incompressible stripe into two incompressible edge channels by employing the self-consistent method.  By varying the magnetic field (hence the average filling factor) the system can undergo a transition from compressible state, bulk incompressible state, to edge incompressible sate. With increasing temperature these incompressible states collapse eventually. The width of edge channels shrink from both sides as the temperature is rising until vanishes, while the bulk state collapses from the middle separated into two edge channels by a density ramp, and the width of these edge channels shrinks at the same time.

%
\bibliographystyle{unsrt}
\bibliography{Decay_of_incompressible_stripes_arxiv.bbl}

\begin{thebibliography}{10}

\bibitem{PRB.46.4026}
D.~B. Chklovskii, B.~I. Shklovskii, and L.~I. Glazman.
\newblock Electrostatics of edge channels.
\newblock {\em Phys. Rev. B}, 46(7):4026--4034, Aug 1992.

\bibitem{PRB.47.12605}
D.~B. Chklovskii, K.~A. Matveev, and B.~I. Shklovskii.
\newblock Ballistic conductance of interacting electrons in the quantum hall
  regime.
\newblock {\em Phys. Rev. B}, 47(19):12605--12617, May 1993.

\bibitem{physicab.298.155}
Z.D Kvon, E.B Olshanetsky, M~Casse, A.Y Plotnikov, D.K Maude, J.C Portal, and
  A.I Toropov.
\newblock Iqhe and fqhe in a wire with incompressible and compressible strips.
\newblock {\em Physica B: Condensed Matter}, 298(1每4):155--158, 2001.

\bibitem{physicae.3.30}
Josef Oswald.
\newblock A new model for the transport regime of the integer quantum hall
  effect: The role of bulk transport in the edge channel picture.
\newblock {\em Physica E: Low-dimensional Systems and Nanostructures},
  3(1每3):30--37, 1998.

\bibitem{physicae.20.43}
S.~Komiyama, O.~Astafiev, and T.~Machida.
\newblock Application of quantum hall edge channels.
\newblock {\em Physica E: Low-dimensional Systems and Nanostructures},
  20(1每2):43--56, 2003.

\bibitem{physicae.40.1217}
A.~Siddiki, D.~Eksi, E.~Cicek, A.I. Mese, S.~Aktas, and T.~Hakioglu.
\newblock Theoretical investigation of the electron velocity in quantum hall
  bars, in the out of linear response regime.
\newblock {\em Physica E: Low-dimensional Systems and Nanostructures},
  40(5):1217--1219, 2008.

\bibitem{physicae.40.1232}
E.V. Deviatov, V.T. Dolgopolov, A.~Lorke, D.~Reuter, and A.D. Wieck.
\newblock Transport across the incompressible strip in the fractional quantum
  hall effect regime.
\newblock {\em Physica E: Low-dimensional Systems and Nanostructures},
  40(5):1232--1234, 2008.

\bibitem{arXiv.1202.6681v1}
Venkatachalam Vivek, Hart Sean, Pfeiffer Loren, West Ken, and Yacoby Amir.
\newblock Local thermometry of neutral modes on the quantum hall edge.
\newblock {\em arXiv:1202.6681}, 2012.

\bibitem{physicae.44.1367}
N.~Boz Yurdaan, K.~Akgungor, A.~Siddiki, and I.~Sokmen.
\newblock Theoretical investigation of local electron temperature in quantum
  hall systems.
\newblock {\em Physica E: Low-dimensional Systems and Nanostructures},
  44(7每8):1367--1371, 2012.

\bibitem{PRB.73.045333}
S.~Komiyama, H.~Sakuma, K.~Ikushima, and K.~Hirakawa.
\newblock Electron temperature of hot spots in quantum hall conductors.
\newblock {\em Phys. Rev. B}, 73:045333, Jan 2006.

\bibitem{physicae.4.79}
G.~Nachtwei.
\newblock Breakdown of the quantum hall effect.
\newblock {\em Physica E: Low-dimensional Systems and Nanostructures},
  4(2):79--101, 1999.

\bibitem{cage}
M.~E. Cage.
\newblock Dependence of quantized hall effect breakdown voltage on magnetic
  field and current.
\newblock {\em J. Res. Natl. Inst. Stand. Technol.}, 98:361, 1993.

\bibitem{PRL.51.1374}
M.~E. Cage, R.~F. Dziuba, B.~F. Field, E.~R. Williams, S.~M. Girvin, A.~C.
  Gossard, D.~C. Tsui, and R.~J. Wagner.
\newblock Dissipation and dynamic nonlinear behavior in the quantum hall
  regime.
\newblock {\em Phys. Rev. Lett.}, 51:1374--1377, Oct 1983.

\bibitem{RMP.80.1083}
Chetan Nayak, Steven~H. Simon, Ady Stern, Michael Freedman, and Sankar
  Das~Sarma.
\newblock Non-abelian anyons and topological quantum computation.
\newblock {\em Rev. Mod. Phys.}, 80:1083--1159, Sep 2008.

\bibitem{Science.318.766}
Markus Konig, Steffen Wiedmann, Christoph Brune, Andreas Roth, Hartmut Buhmann,
  Laurens~W. Molenkamp, Xiao-Liang Qi, and Shou-Cheng Zhang.
\newblock Quantum spin hall insulator state in hgte quantum wells.
\newblock {\em Science}, 318(5851):766--770, 2007.

\bibitem{PRL.101.256802}
K.~Hashimoto, C.~Sohrmann, J.~Wiebe, T.~Inaoka, F.~Meier, Y.~Hirayama, R.~A.
  R\"omer, R.~Wiesendanger, and M.~Morgenstern.
\newblock Quantum hall transition in real space: From localized to extended
  states.
\newblock {\em Phys. Rev. Lett.}, 101:256802, Dec 2008.

\bibitem{Nature.427.328}
S.~Ilani, J.~Martin, E.~Teitelbaum, J.~H. Smet, D.~Mahalu, and A.~Umansky,
  V.~Yacoby.
\newblock {The microscopic nature of localization in the quantum Hall effect}.
\newblock {\em Nature}, 427:328--332, 2004.

\bibitem{PRL.107.176809}
Keji Lai, Worasom Kundhikanjana, Michael~A. Kelly, Zhi-Xun Shen, Javad Shabani,
  and Mansour Shayegan.
\newblock Imaging of coulomb-driven quantum hall edge states.
\newblock {\em Phys. Rev. Lett.}, 107:176809, Oct 2011.

\bibitem{NJOP.14.083015}
M~E Suddards, A~Baumgartner, M~Henini, and C~J Mellor.
\newblock Scanning capacitance imaging of compressible and incompressible
  quantum hall effect edge strips.
\newblock {\em New Journal of Physics}, 14(8):083015, 2012.

\bibitem{PRB.38.4218}
Ulrich Wulf, Vidar Gudmundsson, and Rolf~R. Gerhardts.
\newblock Screening properties of the two-dimensional electron gas in the
  quantum hall regime.
\newblock {\em Phys. Rev. B}, 38(6):4218--4230, Aug 1988.

\bibitem{SSC.67.1019}
A.L. Efros.
\newblock Non-linear screening and the background density of 2deg states in
  magnetic field.
\newblock {\em Solid State Communications}, 67(11):1019 -- 1022, 1988.

\bibitem{RMP.81.109}
A.~H. Castro~Neto, F.~Guinea, N.~M.~R. Peres, K.~S. Novoselov, and A.~K. Geim.
\newblock The electronic properties of graphene.
\newblock {\em Rev. Mod. Phys.}, 81(1):109--162, Jan 2009.

\bibitem{PRB.45.11354}
A.~L. Efros.
\newblock Homogeneous and inhomogeneous states of a two-dimensional electron
  liquid in a strong magnetic field.
\newblock {\em Phys. Rev. B}, 45(19):11354--11357, May 1992.

\bibitem{PRB.50.7757}
Karlheinz Lier and Rolf~R. Gerhardts.
\newblock Self-consistent calculations of edge channels in laterally confined
  two-dimensional electron systems.
\newblock {\em Phys. Rev. B}, 50(11):7757--7767, Sep 1994.

\bibitem{PRB.56.13519}
J.~H. Oh and Rolf~R. Gerhardts.
\newblock Self-consistent thomas-fermi calculation of potential and current
  distributions in a two-dimensional hall bar geometry.
\newblock {\em Phys. Rev. B}, 56(20):13519--13528, Nov 1997.

\bibitem{PRB.68.125315}
A.~Siddiki and Rolf~R. Gerhardts.
\newblock Thomas-fermi-poisson theory of screening for laterally confined and
  unconfined two-dimensional electron systems in strong magnetic fields.
\newblock {\em Phys. Rev. B}, 68(12):125315, Sep 2003.

\bibitem{PRB.70.195335}
Afif Siddiki and Rolf~R. Gerhardts.
\newblock Incompressible strips in dissipative hall bars as origin of quantized
  hall plateaus.
\newblock {\em Phys. Rev. B}, 70(19):195335, Nov 2004.

\bibitem{arXiv.1203.4985}
Ahmet~Emre Kavruk, Teoman Ozturk, Ulfet Atav, and Huseyin Yuksel.
\newblock The effect of the frozen and pinned surface approximations on the
  spatial distribution of incompressible and compressible strips in quantum
  hall regime.
\newblock {\em arXiv:1203.4985}, 2012.

\bibitem{arXiv.1210.8425}
A.~Salman, M.~B. Yucel, and A.~Siddiki.
\newblock Edge electrostatics revisited.
\newblock {\em arXiv:1210.8425v1}, 2012.

\bibitem{arXiv.0906.3796}
A.~Siddiki U.~Erkarslan, G.~Oylumluoglu.
\newblock Edge-to-bulk transition of the iqhe at cleaved edge overgrown
  samples: a screening theory based experimental proposal.
\newblock {\em arXiv:0906.3796}, 2009.

\bibitem{SSC.54.479}
S.~Komiyama, T.~Takamasu, S.~Hiyamizu, and S.~Sasa.
\newblock Breakdown of the quantum hall effect due to electron heating.
\newblock {\em Solid State Communications}, 54(6):479 -- 484, 1985.

\end{thebibliography}

\end{document}